\documentclass[epj, final]{svjour}

\usepackage{amsmath,amstext,amssymb}
\usepackage{graphicx}
\usepackage{dcolumn}
\usepackage{bm}

\newcommand{\V}[1]{{\bm #1}}

\newcommand{\pdiff}[3][]{\frac{\partial^{#1} #2}{\partial #3^{#1}}}

\begin{document}

\title{Quantum kinetic theory of phonon-assisted carrier transitions in nitride-based quantum-dot systems}

\author{J. Seebeck\inst{1} \and T.R. Nielsen\inst{1} \and P. Gartner\inst{1,2} \and F. Jahnke\inst{1}}

\institute{Institute for Theoretical Physics, University of Bremen, 28334 Bremen, Germany \and
           National Institute for Materials Physics, POB MG-7, Bucharest-Magurele, Romania}

\date{\today}

\abstract{
A microscopic theory for the interaction of carriers with LO phonons
is used to study the ultrafast carrier dynamics in nitride-based
semiconductor quantum dots.  It is shown that the efficiency of
scattering processes is directly linked to quasi-particle
renormalizations. The electronic states of the interacting system are
strongly modified by the combined influence of quantum confinement and
polar coupling.  Inherent electrostatic fields, typical for InGaN/GaN
quantum dots, do not limit the fast scattering channels.
}


\maketitle

\section{Introduction}

In the rapidly evolving field of semiconductor quantum dots (QDs),
which provide many new applications in optoelectronics, nitride-based
systems are of growing interest due to their large range of possible
emission frequencies.\cite{Nakamura:00,Cho:02,Kalliakos:04} The
interaction of carriers with LO phonons is an important scattering
process which determines the redistribution of carriers between the
localized QD states and between QD and wetting-layer (WL) states.  Furthermore, these
interaction processes lead to optical dephasing which is of central
importance in coherent optical experiments and determines the
homogeneous broadening of absorption and gain spectra.  In
nitride-based QDs, the role of interaction with LO phonons is
increased due to the intermediate polar coupling strength in this
material system.  On the other hand, for QDs with wurtzite crystal
structure grown along the c-axis, strong built-in fields modify the
electronic states via the quantum-confined Stark
effect.\cite{Bernardini:98} The aim of this paper is to determine the
efficiency of carrier-phonon interaction under these conditions.

While a simplified treatment of carrier-phonon interaction in QDs,
based on Fermi's golden rule, leads to the prediction of a
phonon-bottleneck (see \cite{Benisty:95} and references therein),
recently it has been emphasized that perturbation theory is not
applicable for QD systems and that the interaction needs to be treated
in the polaron picture.\cite{Inoshita:97,Verzelen:02} A
quantum-kinetic theory has been used to describe carrier scattering
processes due to QD polarons.\cite{Seebeck:05} We extend this model by
including also the optical generation of carriers by means of a short laser pulse.
Naturally linked to this is the dephasing of the coherently driven interband polarization, which is described on the same quantum-kinetic level as the carrier scattering.
Furthermore, we study intermediate polar coupling and include a self-consistent description of the quantum-confined Stark effect for nitride-based QDs.

\section{Quantum-dot model}

We consider a system of self-assembled InGaN/GaN QDs, grown on the
c-plane in wurtzite crystal structure where the spontaneous and
piezoelectric polarization along the c-axis cause large electrostatic
fields.  The calculation of single-particle properties for these QDs
is a challenging task on its own.  Considerable recent progress has
been made in $\V{k} \cdot \V{p}$ and tight-binding models for realistic
QD geometries of various material systems to determine the
single-particle wave functions and energies and the resulting
free-particle optical transitions.\cite{Ranjan:03,Bagga:03,Fonoberov:03,Andreev:00} With the investigation of many-body interaction
effects for given single-particle states, this paper follows a
different aim.  Previous calculations of
carrier-carrier\cite{Nielsen:04,Nielsen:05} and
carrier-phonon\cite{Seebeck:05} scattering in QDs showed that
particular details of the wavefunctions are less important, while
results are more sensitive to the energy stucture (level spacing and
degeneracy).

For our purpose, we use a simplified model for the single-particle
states of the QD system based on a separation ansatz
$\Phi(\vec{r})=\varphi(\vec{\rho})~\xi(z)$ for the wave functions.
The in-plane part $\varphi(\vec{\rho})$ describes either the localized
QD states or the delocalized WL states which form a
continuum at higher energies.  In growth direction,
perpendicular to the WL plane, the carrier confinement potential is
modified by intrinsic electrostatic fields.  Due to the
quantum-confined Stark effect, charges become partly separated and a
screening field builds up which reduces the intrinsic electrostatic
field.  Therefore $\xi(z)$ has to be calculated from a self-consistent
solution of the Schr\"odinger and Poisson equation.  For details on
the QD model, the calculation of wavefunctions and material
parameters, see \cite{Nielsen:05}.

\begin{figure}
   \includegraphics[width=.9\columnwidth]{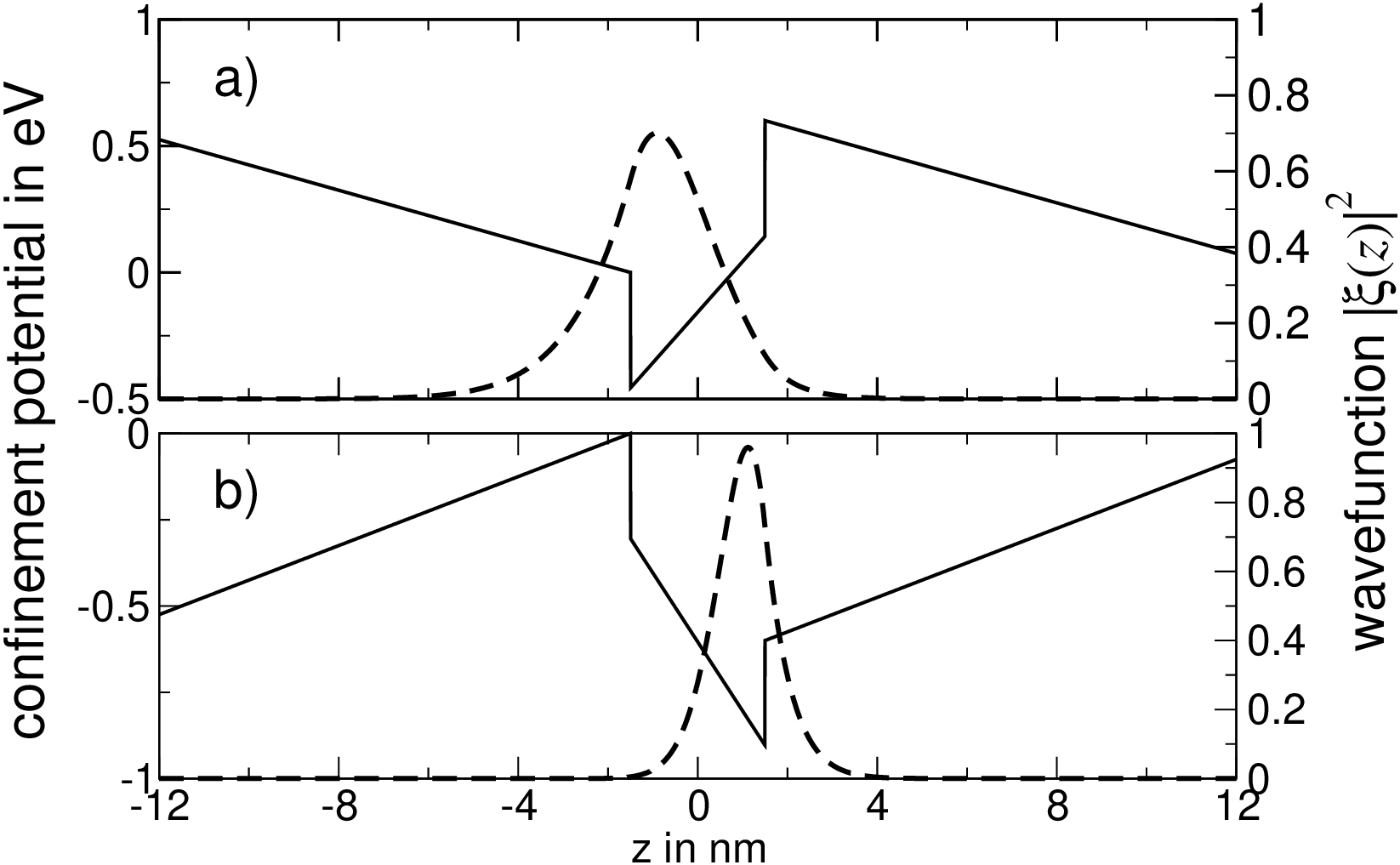}
   \caption{Confinement potential $U(z)$ (solid line) and wave function
            $|\xi(z)|^2$ (dashed line) for electrons (a) and holes (b)
            in growth direction for an electrostatic field of
            2 MV/cm.\label{fig1}}
\end{figure}

We consider QDs with two confined shells which due to their
angular momentum properties are referred to as the s-shell (ground
state) and the p-shell (excited state).  The level spacing between s-
and p-states is 1.2 $\hbar\omega_\text{LO}$ for electrons and 0.4
$\hbar\omega_\text{LO}$ for holes.  For the intrinsic electrostatic
field, we consider a typical field strength of 2 MV/cm.

\section{Many-body theory}

A perturbative treatment of the carrier interaction with LO phonons
is only possible when modifications of the single-particle
states and energies due to the polar coupling remain small.
In QDs the localized states with discrete energies are subject to
hybridization effects even in materials with weak polar coupling.\cite{Inoshita:97,Verzelen:02,Seebeck:05}
The QD polaron states are broadened due to the interaction with carriers in the WL
continuum, which ensures dissipation.\cite{Seebeck:05} Quasi-particle renormalizations can
be described using the retarded Green's function, which obeys
the Dyson equation
\begin{align}
   & \left[ i\hbar\pdiff{}{\tau}{}-\epsilon^{\text{c}}_\alpha\right]
   G^{\text{c},\text{R}}_\alpha(\tau) = \delta(\tau) \nonumber\\
&+ \int_0^\tau d\tau'
   \sum_\beta G^{\text{c},\text{R}}_\beta(\tau-\tau')
 ~D^{\text{cc},
   >}_{\alpha\beta}(\tau-\tau')~G^{\text{c},\text{R}}_\alpha(\tau')
   ~,
\end{align}
for electrons in the conduction band and a similar equation with
$D^{\text{vv},<}$ for the valence band.
This equation describes the polaron in the self-consistent, diagonal,
random-phase approximation (RPA) \cite{Kral:98,Seebeck:05}.
The single-particle energies for either localized or
extended states are given by $\epsilon^{\lambda}_\alpha$ , with $\; \lambda =
\text{c,v}$ being the band index.
Here, the corresponding wave functions (which incorporate effects of the built-in
field) enter via the interaction matrix elements $M^{\lambda}_{\alpha,\beta}$.
They are included in the phonon propagator
\begin{align}
   D^{\lambda \mu,\gtrless}_{\alpha\beta}(\tau)
 &= \sum_{\vec{q}} ~ M^{\lambda}_{\alpha\beta}(\vec{q})
 M^{\mu}_{\beta\alpha}(-\vec{q})\nonumber\\
 &\times ~\left[~N_\text{LO} ~e^{\pm i\omega_\text{LO}\tau}
        + (1+N_\text{LO}) ~e^{\mp i\omega_\text{LO}\tau}\right] ~,
\end{align}
where $N_\text{LO}$ is a Bose-Einstein function for the population of
the phonon modes (assumed to act as a bath in thermal equilibrium) and
$\omega_\text{LO}$ is the LO-phonon frequency.  The Fourier transform
of the retarded GF directly provides the spectral function
$\widehat{G}_\alpha(\omega)=-2~\text{Im}~G^\text{R}_\alpha(\omega)$
which reflects the density of states (DOS) for a carrier in state $\alpha$.

\begin{figure}
   \includegraphics[width=.9\columnwidth]{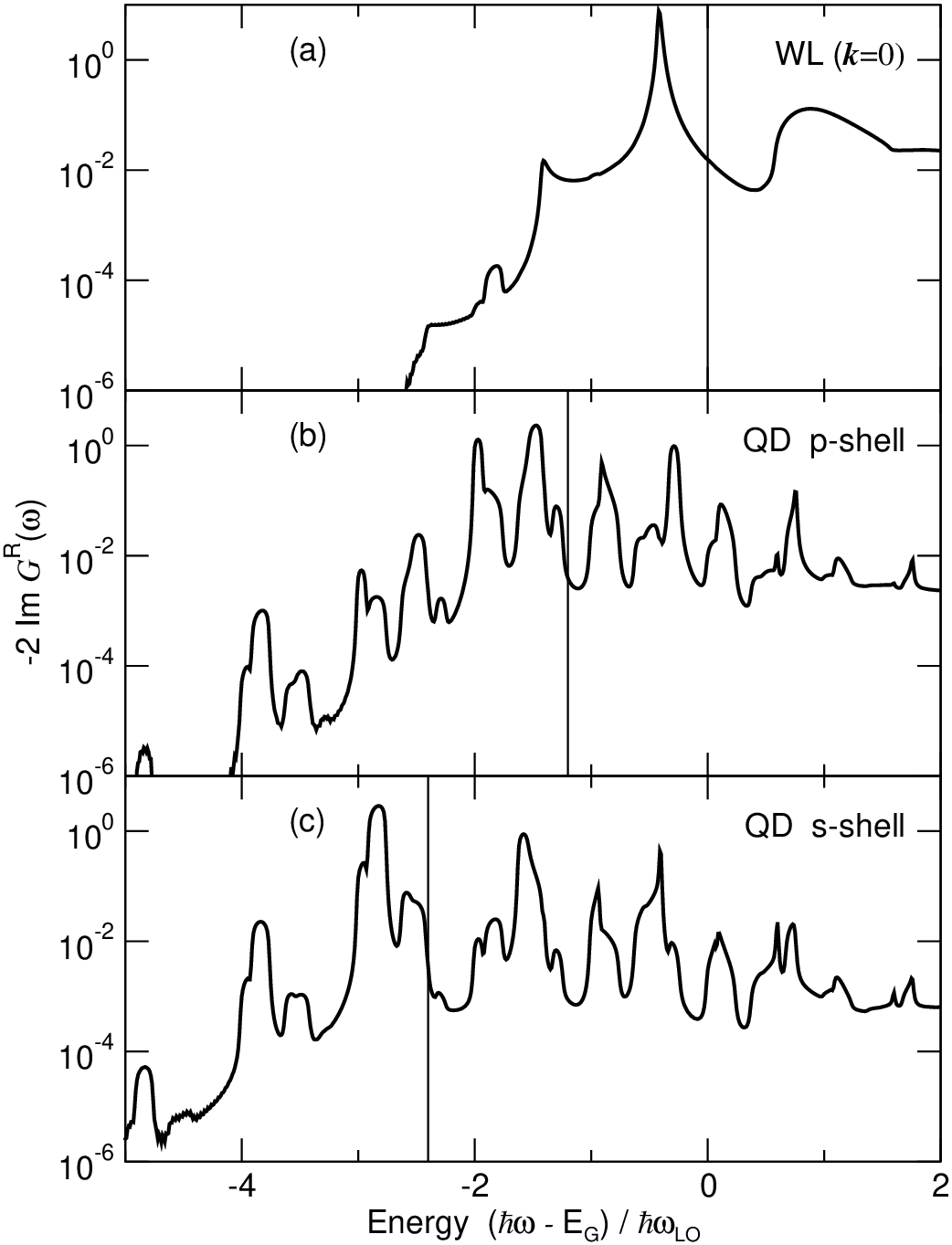}
   \caption{Spectral function for a WL electron with zero momentum (a)
            and for a QD electron in the p-shell (b) and in the
            s-shell (c). Vertical lines indicate the positions of the
            $\delta$-functions for a non-interacting carriers.
           \label{fig2}}
\end{figure}

For non-interacting carriers, the spectral function contains
$\delta$-functions at the free-particle energies, reflecting the
infinite lifetime of these states (if recombination on a much
longer timescale is neglected).  The spectral functions for carriers in
the QD-WL system interacting with lattice distortions are shown in
Fig.~\ref{fig2}. For a WL electron with zero momentum (at the band
edge), the $\delta$-like spectral function in the absence of
interaction (indicated by a vertical line in Fig.~\ref{fig2}a) is
broadened. The corresponding finite lifetime of the state is due to
possible scattering events which are self-consistently calculated in
terms of renormalized states and in arbitrary order in the interaction
by means of a direct numerical solution of Eqs.~(1) and (2). Furthermore,
the energy lowering of the interacting state (polaron shift) and weak
LO phonon satellites are obtained.

For the discrete QD states, Figs.~\ref{fig2}b) and c) show that the free-particle picture is strongly modified.
On this level, a direct comparison between the the weak-polar coupling material system InGaAs (Fig. 1 of Ref. \cite{Seebeck:05}) and the InGaN system with intermediate polar coupling is possible.
In the first case, one can clearly distinguish between phonon replicas of the QD states which are spaced by the LO-phonon energy $\hbar\omega_\text{LO}$ around the central peak, and the hybridization effect which leads to a splitting of these peaks that is smaller than the LO-phonon energy (when the detuning between the QD level spacing and the LO-phonon energy is small).
In the InGaN material system, the hybridization is much stronger and one can no longer distinguish between phonon replicas and their splitting.
Moreover, the obtained rich multi-peak structure provides a stronger overlap between the spectral functions which increases the scattering efficiency discussed below.
Broadening of the resonances is provided by the coupling to the WL states and the finite LO-phonon lifetime (assumed to be 5 ps).

In the markovian limit, the transition rate between two states
$\alpha$ and $\beta$ is proportional to the overlap
$\Lambda_{\alpha\beta}=\int d\omega \; \widehat{G}_\alpha(\omega) \;
\widehat{G}_\beta(\omega\pm\omega_\text{LO})$.  In the free-particle
picture, considering the $\delta$-like DOS, the results of Fermi's
golden rule are recovered.  The level-spacing has to match exactly the
LO-phonon energy $\hbar\omega_\text{LO}$, otherwise the overlap
vanishes and carrier transitions are not possible.  In the polaron
picture, strong overlap between the spectral functions of the QD
states as well as between the QD and WL states is obtained which already
indicates the possibility of efficient scattering processes. On
ultrafast timescales, the Markov approximation is not valid and
quantum-kinetic effects additionally increase the scattering
rates.

To study the carrier relaxation from p-shell into s-shell, we
consider the optical excitation of carriers with a short (100 fs)
laser pulse having spectral overlap only with the p-shell.
The
coupled carrier and polarization dynamics are described by the kinetic
equation for the single-particle density matrix
${\rho}_\alpha(t)$, which contains optically induced
interband transition amplitudes $\psi_\alpha$ and occupation
probabilities $f_\alpha^{\text{c,v}}$ according to \cite{Haug_Jauho}
\begin{align}\label{eq:gk}
   \pdiff{}{t}\rho_\alpha(t) &=\pdiff{}{t}\begin{pmatrix} f_{\alpha}^\text{c}
  & \psi_{\alpha} \\ \psi_{\alpha}^* & f_{\alpha}^\text{v}
  \end{pmatrix}=\nonumber \\
  &  = \left[ \begin{pmatrix} \tilde\epsilon^\text{c}_\alpha
  & -\Omega_\alpha \\ -\Omega_\alpha^*
  & \tilde\epsilon^\text{v}_\alpha \end{pmatrix},\rho_\alpha \right]
  + \pdiff{}{t} \rho_\alpha(t)\Big|_\text{coll.} ~.
\end{align}
In this matrix notation of the semiconductor Bloch equations,
Hartree-Fock renormalizations due to the carrier Coulomb interaction
are included in the single-particle
energies $\tilde\epsilon^\lambda_\alpha=\epsilon^\lambda_\alpha+\sum_\beta v^{\lambda\lambda}_{\alpha\beta} (f^\lambda_\alpha-\delta_{\lambda\text{v}})$ and in the Rabi energy
$\Omega_\alpha=\vec{d}\vec{E}(t)+\sum_\beta v^\text{cv}_{\alpha\beta}\psi_\alpha$.
Here $\vec{d}$ is the dipole coupling, $\vec{E}(t)$ the laser pulse and $v^{\lambda\lambda'}_{\alpha\beta}$ are the matrix elements of the Coulomb interaction \cite{Nielsen:05}.
The quantum kinetic description of the carrier-phonon interaction, based on the generalized Kadanoff-Baym ansatz
\cite{Haug_Jauho},
\begin{align}\label{eq:gk_coll}
   \pdiff{}{t}&\rho^{\lambda\lambda' }_\alpha(t)\Big|_\text{coll.}
  = \int_{-\infty}^{t} dt' ~\sum_{\beta\mu} \nonumber\\
  &\times \Big\{
  G^{\lambda,\text{R}}_\beta(t,t')~\left[G^{\lambda',\text{R}}_\alpha(t,t')
  \right]^* \nonumber\\
  & ~~~~~    \times\Big[
  -~
  [\delta_{\lambda\mu}-\rho^{\lambda\mu}_\beta(t')]
  ~\rho^{\mu\lambda'}_\alpha(t')~
  D^{\lambda\mu,>}_{\alpha\beta}(t,t')       \nonumber\\
  & ~~~~~       ~~~~~    +~
  \rho^{\lambda\mu}_\beta(t')~
  [\delta_{\mu\lambda'}-\rho^{\mu\lambda'}_\alpha(t')]~
  D^{\lambda\mu,<}_{\alpha\beta}(t,t') \Big] \nonumber\\
  & +
  G^{\lambda,\text{R}}_\alpha(t,t')~\left[G^{\lambda',\text{R}}_\beta(t,t')
  \right]^* \nonumber\\
  & ~~~    \times\Big[ +~ [\delta_{\lambda\mu}-\rho^{\lambda\mu}_\alpha(t')]~
  \rho^{\mu\lambda'}_\beta(t')
  ~D^{\mu\lambda',>}_{\alpha\beta}(t,t') \nonumber\\
  & ~~~       ~~~~~    -~ \rho^{\lambda\mu}_\alpha(t')~
  [\delta_{\mu\lambda'}-\rho^{\mu\lambda'}_\beta(t')]
  ~D^{\mu\lambda',<}_{\alpha\beta}(t,t') \Big] \Big\} ~,
\end{align}
represents a generalization of the formulation in
Ref.~\cite{Seebeck:05} to the coupled carrier and polarization
dynamics and describes scattering, dephasing, and energy
renormalization contributions.
The RPA self-energy was used, and Hartree terms were neglected, since we consider only low-density populations.
In the markovian limit, the slowly varying part of $\rho(t')$ is taken out of the integral and memory kernels of the type
$\Lambda^{\lambda\mu}_{\alpha\beta}=\int~dt'
G^{\lambda,\text{R}}_\alpha(t') \big[G^{\mu,\text{R}}_\beta(t')\big]^*
e^{\pm i\omega_\text{LO}t'}$
determine the scattering efficiency between two states $\alpha$ and
$\beta$ according to the overlap of spectral functions in the
frequency domain, as stated above.  In the non-markovian case, the
memory kernel contains also the single particle density matrix.  This gives
rise to memory effects since the actual time evolution depends
explicitly on the history of the system.

\begin{figure}
   \includegraphics[width=.9\columnwidth]{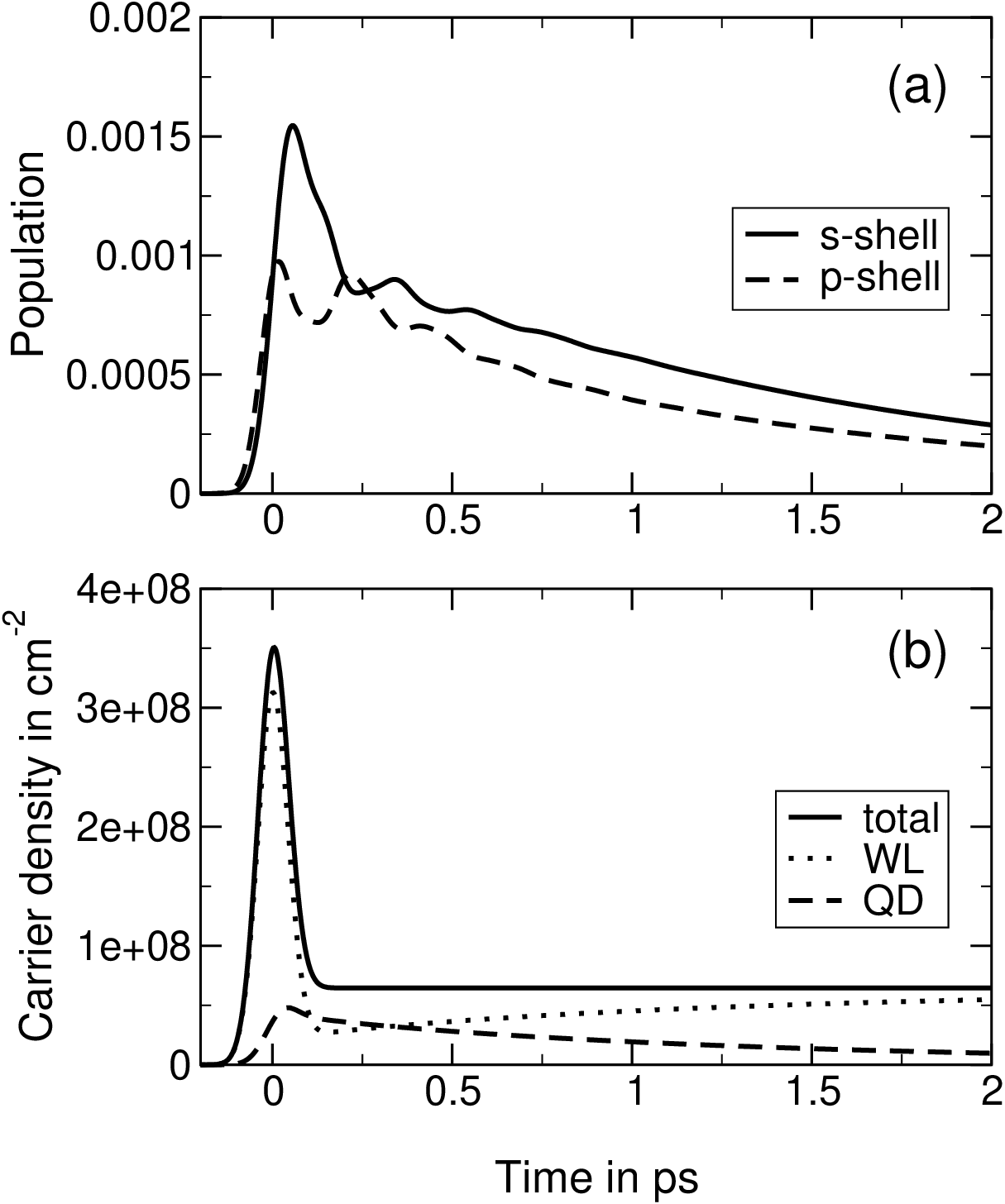}
   \caption{(a) Time evolution of the electron population in the
   p-shell (dashed line) and s-shell (solid line) after optical
   excitation with a 100 fs pulse resonant with the
   p-shell.(b) Total carrier density (solid line), and carrier
   density in the WL (dotted line) and QD (dashed line) versus time.
   \label{fig3}}
\end{figure}

The considered level-spacing of 1.2 $\hbar\omega_\text{LO}$ for electrons and 0.4 $\hbar\omega_\text{LO}$ for holes provides no carrier transitions in a free-particle picture.
Results of the quantum-kinetic calculation are shown in Fig.~\ref{fig3}.
Due to pulse excitation, first the p-shell is populated.
The carrier relaxation is so fast that already during the optical pulse a strong redistribution from the p-shell to the s-shell takes place.
The ultrafast scattering processes are connected with strong dephasing of the coherent interband polarization, which is included on the same quantum-kinetic level in Eq. (4).
The strong dephasing leads to a time evolution of the interband polarization, which practically follows the optical pulse (not shown).
Correspondingly, the coherent regime is limited to the duration of the 100 fs pump pulse.
A transient strong increase of the total carrier density is obtained during the coherent regime, as can be seen in Fig. 3 b).

The time-domain oscillations in Fig. 3a) indicate the strong coupling between carriers and phonons.
The oscillation frequency is connected to the hybridization in the DOS.
The phenomenon is analogous to Rabi-oscillations, though not due to coupling of carriers to the optical field, as recently discussed for QDs in Refs. \cite{Foerstner:03,Wang:05}, but rather due to coupling of carriers to the bosonic field of the phonons.
Note that the oscillations appear after the optical pump pulse in the incoherent regime.
Hence they are not connected to optical coherence.
At larger times, the possibility of carrier transitions between QD and WL states leads to a redistribution of carriers from the QD states into the WL.
The corresponding reduction of the QD population in connection with the increasing WL carrier density can be seen in Figs. 3 a) and b), respectively.
Of course, the total carrier density remains constant after the pulse.

The ultrafast scattering and dephasing processes are the direct result of the polaronic renormalizations.
Only within a non-markovian calculation the above mentioned oscillations in connection with the strong-coupling regime for the carrier-phonon interaction can be obtained.\cite{Seebeck:05}
In this respect, there is no qualitative difference between the InGaAs and InGaN material systems.
In the latter case, the larger polar coupling increases the scattering efficiency while the electron and hole charge separation induced by the built-in fields reduces only inter-band but not intra-band processes.

In the presented example, a weak excitation pulse has been used to
justify the restriction of the interaction to LO phonons. At elevated
excitation densities, carrier-carrier scattering additionally
contributes to fast scattering processes.\cite{Nielsen:05}

\section{Conclusion}

It has been shown that in nitride-based QD systems with intermediate
polar coupling strong modifications of the free-particle states are
obtained due to the interaction of carriers with LO phonons.
Therefore, Fermi's golden rule is not applicable and a quantum-kinetic
description is necessary, which provides ultrafast scattering channels.

This work was supported by the Deutsche Forschungsgemeinschaft.  We
acknowledge a grant for CPU time at the NIC, Forschungszentrum J\"ulich.


\end{document}